\documentclass{article} %

\usepackage{mathptmx}

\usepackage{graphicx}
\usepackage{pifont}
\usepackage{latexsym}
\usepackage{amsfonts}
\usepackage{amsmath}
\usepackage{amssymb}
\usepackage{float}
\usepackage{subfig} %

\def\shortcite{\cite}

\newcommand{\p}{{\rm P}}
\newcommand{\np}{{\rm NP}}
\newcommand{\npc}{\np\text{-complete}}
\newcommand{\elec}{\ensuremath{\cal E}}
\newcommand{\thetatwo}{\ensuremath{\Theta_2^p}}

\newtheorem{theorem}{Theorem}

\newtheorem{lemma}[theorem]{Lemma}
\newcommand\qedblob{\ding{113}}
\def\literalqed{{\ \nolinebreak\hfill\mbox{\qedblob\quad}}}

\newtheorem{example}{Example}
\newtheorem{definition}{Definition}
\newenvironment{proofs}{\noindent{\bf Proof.}\hspace*{1em}}{\literalqed\bigskip}

\showhyphens{}

\newcommand{\score}[1]{\ensuremath{{\rm score}(#1)}}
\newcommand{\scoresub}[2]{\ensuremath{{\rm score}_{#1}(#2)}}
\newcommand{\prob}[3]{
\noindent
{\textbf{Name:}} #1

\noindent
{\textbf{Given:}} #2

\noindent
{\textbf{Question:}} #3}

\begin{document}
\sloppy

\title{Modeling Single-Peakedness for Votes with Ties}

\author{
Zack Fitzsimmons \\
  College of Computing and Inf.\ Sciences\\
  Rochester Inst.\ of Technology \\
  Rochester, NY 14623 \and
  Edith Hemaspaandra \\
  Dept.~of Computer Science\\
  Rochester Inst.\ of Technology \\
  Rochester, NY 14623}%

\date{June 18, 2016}

\maketitle

\begin{abstract}
Single-peakedness is one of the most important and well-known domain restrictions on preferences.
The computational study of single-peaked electorates has largely been restricted to elections with tie-free votes, and
recent work that studies the computational complexity of manipulative attacks for single-peaked elections for votes with
ties has been restricted to nonstandard models of single-peaked preferences for top orders.  We study the
computational complexity of manipulation for votes with ties for the standard model of single-peaked preferences and for single-plateaued preferences.
We show that these models avoid the anomalous complexity behavior exhibited by the other models. %
We also state a surprising result on the relation between the societal axis and the complexity of manipulation for
single-peaked preferences.
\end{abstract}

\section{Introduction}\label{sec:intro}

Elections are a general and widely used framework for preference aggregation in human and
artificial intelligence applications. An important negative result from social choice
theory, the
Gibbard-Satterthwaite Theorem~\cite{gib:j:polsci:manipulation,sat:j:polsci:manipulation},
states that every reasonable election system is manipulable. However, even though every election system is
manipulable, it may be computationally infeasible to determine how to manipulate the outcome.

Bartholdi, Tovey, and Trick~\shortcite{bar-tov-tri:j:manipulating} introduced the computational study of the manipulation problem
and this began an exciting line of research that explores the computational complexity
of different manipulative attacks on elections (see, e.g.,~\cite{fal-hem-hem:j:cacm-survey}).

The notion of single-peaked preferences introduced by
Black~\shortcite{bla:j:rationale-of-group-decision-making} is the most important restriction on
preferences from political science and economics and is naturally an important case to consider computationally.
Single-peakedness models the preferences of a collection of voters with respect to a given axis (a total ordering
of the candidates). Each voter in a single-peaked election has a single most-preferred
candidate (peak) on the axis and the farther candidates are to the left/right from her peak the less preferred they are. %
Single-plateauedness extends this to model when each voter has multiple most-preferred candidates that
appear sequentially on the axis, but are otherwise single-peaked~\cite{bla:b:polsci:committees-elections}.

This standard model of single-peaked preferences has many desirable social-choice properties.
When the voters in an election are single-peaked, the majority relation is
transitive~\cite{bla:j:rationale-of-group-decision-making}
and there exist voting rules that are strategy-proof~\cite{mou:j:strategy-proof}
(i.e., a voter cannot misrepresent her preferences to achieve a personally better outcome).
Single-peakedness for total orders can also have an effect on 
the complexity of many different election attack problems when compared
to the general case.
The complexity of manipulative attacks often decreases when the voters in an election are
single-peaked~\cite{fal-hem-hem-rot:j:single-peaked-preferences,bra-bri-hem-hem:j:sp2}, and the winner problems
for Kemeny, Dodgson, and Young elections are in \p\ when they are \thetatwo-complete in
general~\cite{hem-hem-rot:j:dodgson,hem-spa-vog:j:kemeny,rot-spa-vog:j:young}.

Most of the abovementioned research on the computational complexity of manipulation of
elections, both for the general case and for single-peaked electorates, has been limited to the
assumption that voters have tie-free votes.
In many real-world scenarios voters have votes with ties, and this
is seen in the online repository {\sc PrefLib}~\cite{mat-wal:c:preflib} that contains several
different preference datasets that contain ties. There are also election systems
defined for votes with ties, e.g., the Kemeny rule and the Schulze rule~\cite{kem:j:no-numbers,sch:j:clone-independent-new}.

Recent work considers the complexity of manipulation for top-order votes 
(votes where all of the ties are between candidates ranked last)~\cite{nar-wal:c:partial-vote-manipulation,men-lar:t:manipulation-partial}.
Fitzsimmons and Hemaspaandra~\shortcite{fit-hem:c:voting-with-ties}
considered the complexity of manipulation, control, and bribery for more general
votes with ties, %
and also the case of manipulation for a nonstandard model of single-peakedness for top-order votes. %
Menon and Larson~\shortcite{men-lar:c:sp-manipulation-partial} later examined the complexity of manipulation and bribery for an equivalent (for top orders)
model of single-peakedness. %

Fitzsimmons and Hemaspaandra~\shortcite{fit-hem:c:voting-with-ties} use the model of possibly single-peaked
preferences from Lackner~\shortcite{lac:c:incomplete-sp-aaai} where a preference profile of votes with ties is
said to be single-peaked with respect to an axis if the votes can be extended to tie-free votes that are
single-peaked with respect to the same axis. Menon and Larson~\shortcite{men-lar:c:sp-manipulation-partial} use
a similar model for top orders that they state is essentially the model of
single-peaked preferences with outside options~\cite{can:j:single-peaked-outside-option}.
Both Fitzsimmons and Hemaspaandra~\shortcite{fit-hem:c:voting-with-ties} and
Menon and Larson~\shortcite{men-lar:c:sp-manipulation-partial} find that these notions of single-peakedness
exhibit anomalous computational behavior where the complexity of manipulation can increase when
compared with the case of single-peaked total orders.

We are the first to study the computational complexity of manipulation for the standard model
of single-peaked preferences for votes with ties,
and for single-plateaued preferences for votes with ties.
In contrast to the recent related work using other %
models of single-peakedness with ties, we
find that the complexity of weighted manipulation for $m$-candidate scoring rules
and for $m$-candidate Copeland$^\alpha$ elections for all $0 \le \alpha < 1$
does not increase when compared
to the cases of single-peaked total orders,
and that the complexity of weighted manipulation does not increase with respect
to the general case of elimination veto elections.
We also compare the social choice properties of these different models, and state a surprising result on the relation
between the societal axis and the complexity of manipulation for single-peaked preferences.

\section{Preliminaries}\label{sec:prelim}

An {\em election} consists of a finite set of candidates $C$ and
a finite collection of voters $V$. We will sometimes refer to this
collection of voters as a {\em preference profile}. An {\em election system}
\elec\ is a mapping from an election to a set of winners, which can
be any subset of the candidate set (the nonunique winner model, our standard model), or at most a
single candidate (the unique winner model).

Each voter in an election
has a corresponding vote (or {\em preference order}) over the set of candidates. 
This is often assumed to be a {\em total order}, i.e., a strict ordering of the candidates
from most to least preferred.
Formally, a total order is a complete, reflexive, transitive,
and antisymmetric binary relation. %
We use ``$>$'' to denote strict preference between two candidates. 

Similarly, a {\em weak order} is a total order without antisymmetry, so each voter can
rank candidates as tied (which we will sometimes refer to as indifference) as long as their
ties are transitive. We use ``$\sim$'' to denote a tie between two candidates. 
A {\em bottom order} is a weak order where all ties are between top-ranked candidates
and a {\em top order} is a weak order where all ties are between bottom-ranked candidates.
Notice that a total order is a weak order, a top order, and a bottom order, that a top order is
a weak order, and that a bottom order is a weak order. Throughout this paper we
will sometimes refer to weak orders as votes with ties.

For some of our results we consider weighted elections, where each voter has an
associated positive integral weight and a voter with weight $w$ counts as $w$ unweighted
voters all voting the same.

\subsection{Election System Definitions}

Our election systems include scoring rules, elimination veto, and Copeland$^\alpha$. We define
each below and the %
extensions we use %
to properly consider votes with ties.
Given an election with $m$ candidates, a scoring rule assigns scores
to the candidates using its corresponding $m$-candidate scoring vector of the
form
$\alpha = \langle \alpha_1, \alpha_2, \dots,
\alpha_m \rangle$ where $\alpha_1 \ge \alpha_2 \ge \dots \ge \alpha_m$ and
each $\alpha_i \in \mathbb{N}$.
So, when the preferences of a voter are a total order, the candidate ranked in position $i$ receives
a score of $\alpha_i$ from that voter. Below we present %
examples of scoring rules and their
corresponding $m$-candidate scoring vector.

\noindent
\textbf{Plurality:} with scoring vector $\langle 1, 0, \dots, 0 \rangle$.
 
\noindent
\textbf{Veto:} with scoring vector $\langle 1, 1, \dots, 1, 0 \rangle$.

\noindent
\textbf{Borda:} with scoring vector $\langle m-1, m-2, \dots, 1, 0 \rangle$.

\noindent
\textbf{Triviality:} with scoring vector $\langle 0, 0, \dots, 0 \rangle$.

To use a scoring rule to determine the outcome of an election containing votes with ties
we must extend the above definition of scoring rules.  We use the definitions of scoring-rule extensions
for weak orders from our previous work~\cite{fit-hem:c:voting-with-ties}, which generalize
the extensions introduced for top orders from Baumeister et al.~\shortcite{bau-fal-lan-rot:c:lazy-voters}
and from Narodytska and Walsh~\shortcite{nar-wal:c:partial-vote-manipulation} which in turn generalizes the %
extensions used by Emerson~\shortcite{eme:j:partial-borda} for Borda. %

Given a weak-order vote, we can write it as $G_1 > G_2 > \cdots > G_r$, where each
$G_i$ is a set of tied candidates, (so in the case of a total order vote each $G_i$ is a singleton).
For each $G_i$, let $k_i = \sum_{j=1}^{i-1} \|G_j\|$ be the number of candidates strictly preferred to %
the candidates in $G_i$. %
We now state the definitions of each of the four extensions. %
In Example~\ref{ex:scoring-rule-ext} we
present an example of how a given weak-order vote is scored using Borda and
each of the scoring-rule extensions.

\noindent
\textbf{Min:} each candidate in $G_i$ receives a score of $\alpha_{k_i + \|G_i\|}$.

\noindent
\textbf{Max:} each candidate in $G_i$ receives a score of $\alpha_{k_i + 1}$.

\noindent
\textbf{Round down:} each candidate in $G_i$ receives a score of $\alpha_{m-r+i}$.

\noindent
\textbf{Average:} each candidate in $G_i$ receives a score of
    $(\sum_{j=k_i + 1}^{k_i + \|G_i\|} \alpha_j)/\|G_i\|.$

For top orders the scoring-rule extensions min, round down, and average
are the same as round up, round down, and average used in the work by
Menon and Larson~\shortcite{men-lar:c:sp-manipulation-partial}.

\begin{example}\label{ex:scoring-rule-ext}
  \normalfont
  Given the candidate set $\{a,b,c,d,e\}$ and the weak order vote $(a \sim b > c \sim d > e)$
  we show the scores assigned to each candidate using Borda and each of our extensions.
We can write the vote as $\{a,b\} > \{c,d\} > \{e\}$, so $G_1 = \{a,b\}$, $G_2 = \{c,d\}$, and
  $G_3 = \{e\}$, and $k_1 = 0$, $k_2 = 2$, and $k_3 = 4$.

  Recall that for total orders, the scoring vector for 5-candidate Borda is $\langle 4,3,2,1,0 \rangle$.

\noindent
\textbf{Borda using min:} $\score{a} = \score{b} = 3$, $\score{c} = \score{d} = 1$, and $\score{e} = 0$.

\noindent
\textbf{Borda using max:} $\score{a} = \score{b} = 4$, $\score{c} = \score{d} = 2$, and $\score{e} = 0$.

\noindent
\textbf{Borda using round down:} $\score{a} = \score{b} = 2$, $\score{c} = \score{d} = 1$, and $\score{e} = 0$.

\noindent
\textbf{Borda using average:} $\score{a} = \score{b} = 3.5$, $\score{c} = \score{d} = 1.5$, and $\score{e} = 0$.

\end{example}

For elimination veto for total
orders, the veto scoring rule is used, %
the candidate with the lowest score is eliminated, and the
rule is repeated on the remaining votes restricted to the remaining candidates until there is one
candidate left~\cite{col-tea:cNEW:manipulation}. We break ties %
lexicographically, and for comparison with related work
our results for elimination veto use the unique winner model,
and for votes with ties we use the min extension. %

Pairwise election systems are one of the most natural cases for considering votes with ties.
Copeland$^\alpha$ is an important and well-known election system %
that is defined using pairwise comparisons between candidates.
In a Copeland$^\alpha$ election each candidate receives one point for each pairwise majority election
with each other candidate she wins and $\alpha$ points for each tie (where $\alpha \in \mathbb{Q}$ and
$0 \le \alpha \le 1$).
For votes with ties we follow the obvious extension also used by
Baumeister et al.~\shortcite{bau-fal-lan-rot:c:lazy-voters} and
Narodytska and Walsh~\shortcite{nar-wal:c:partial-vote-manipulation}.

For Copeland$^\alpha$ elections it will sometimes be easier to refer to the {\em induced majority graph}
of an election. Given an election $(C,V)$ its induced majority graph is constructed as follows.
Each vertex in the induced majority graph corresponds to a candidate in $C$, and for all candidates $a,b \in C$
if $a > b$ by majority then there is an edge from $a$ to $b$ in the induced majority graph. %
We also will refer to the {\em weighted majority graph} of an election, where each edge from $a$ to $b$
in the induced majority graph is labeled with the difference between the number of voters that state $a > b$
and the number of voters that state $b > a$.

\subsection{Election Problems}

The computational study of the manipulation of elections was introduced by
Bartholdi, Tovey, and Trick~\shortcite{bar-tov-tri:j:manipulating}, and %
Conitzer, Sandholm, and Lang~\shortcite{con-lan-san:j:when-hard-to-manipulate} extended this to the
case for weighted voters and a coalition of manipulators. %
We define the constructive weighted coalitional manipulation (CWCM) problem below.

\prob{$\elec$-CWCM}{a set of candidates $C$, a collection of nonmanipulative voters $S$, %
 a collection of manipulative voters $T$,  and a preferred candidate
$p \in C$.}{Does there exist a way to set the votes of $T$ such that $p$ is a winner of
$(C,S \cup T)$ under election system $\elec$?}

For the case of CWCM %
for each of our models of single-peaked preferences we follow the
model introduced by Walsh~\shortcite{wal:c:uncertainty-in-preference-elicitation-aggregation} where the societal axis is given as part of the input to the problem
and the manipulators must state votes that are single-peaked with respect to this
axis (for the corresponding model of single-peakedness). %
(See Section~\ref{sec:variants} for all of the definitions of single-peaked preferences that we use.)

\subsection{Computational Complexity}

For our NP-completeness results we will use the following well-known NP-complete problem.

\prob{Partition}{Given a collection of $t$ positive integers $k_1, \ldots, k_t$ such that $\sum_{i=1}^{t} k_i = 2K$.}{Does there exist a partition of $k_1, \ldots, k_t$ into two subcollections $A$ and $B$ such that
$\sum A = \sum B = K$?} %

\section{Models of Single-Peaked Preferences}\label{sec:variants}

\begin{figure}[h]
  \centering
  \includegraphics{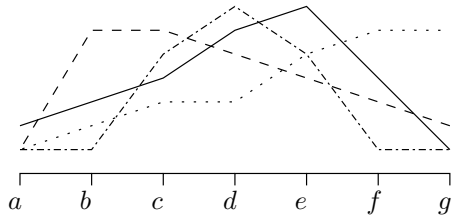}
  \caption{Given the axis $L = a < b < c < d < e < f < g$, the solid line is the single-peaked
  order $(e > d > c \sim f > b > a > g)$, the dashed line is the single-plateaued order
  $(b \sim c > d > e > f >g > a)$, 
the dashed-dotted line is the single-peaked with
outside options order $(d > c \sim e > a \sim b \sim f \sim g)$,
and the dotted line is the possibly single-peaked order
$(f \sim g > e > d \sim c > b > a)$. %
 See Figure~\ref{fig:relation} for the relationships
between the four models.}
  \label{fig:ex-peak}
\end{figure}

We consider %
four important models of single-peaked preferences for
votes with ties. For the following definitions, for a given axis $L$ (a total ordering of the
candidates) and a given preference order $v$, we say that $v$ is strictly increasing (decreasing) along a 
segment of $L$ if each candidate is preferred to the candidate on its left (right) with respect to $L$.
In Figure~\ref{fig:ex-peak} we present an example of each of the four models of single-peaked preferences, and in Figure~\ref{fig:relation} we show how the four models relate
to each other. %
We now give the definition of the standard model of single-peakedness from Black~\shortcite{bla:j:rationale-of-group-decision-making}.

\begin{definition}\label{def:standard}
Given a preference profile $V$ of weak orders over a set of candidates $C$,
$V$ is single-peaked with respect to a total ordering of the candidates $L$ (an axis)
if for each voter $v \in V$, $L$ can be split into three segments $X$, $Y$, and $Z$ ($X$ and $Z$ can
each be empty) such that $Y$ contains only the most preferred candidate of $v$, $v$ is strictly
increasing along $X$ and $v$ is strictly decreasing along $Z$.
\end{definition}

Observe that for a preference profile of votes with ties to be single-peaked with respect to an axis,
each voter can have a tie between at most two candidates at each position in her preference order
since the candidates must each appear on separate sides of her %
peak. Otherwise the preference order
would not be {\em strictly} increasing/decreasing along the given axis.

The model of single-plateaued preferences
extends single-peakedness by allowing
voters to have multiple most preferred candidates
(an indifference plateau)~\cite{bla:b:polsci:committees-elections}. This is defined by extending
Definition~\ref{def:standard} so that $Y$ can contain multiple candidates.
Lackner~\shortcite{lac:c:incomplete-sp-aaai} recently introduced another extension to single-peaked preferences,
which we refer to as ``possibly single-peaked preferences''
throughout this paper. %
A preference profile is possibly single-peaked with respect to a given axis if there
exists an extension of each preference order to a total order such that the new preference profile
of total orders is single-peaked. This can be stated without referring to extensions,
for votes with ties,
in the following way.

\begin{definition}
Given a preference profile $V$ of weak orders over a set of candidates $C$,
$V$ is possibly single-peaked with respect to a total ordering of the candidates $L$ (an axis)
if for each voter $v \in V$, $L$ can be split into three segments $X$, $Y$, and $Z$ ($X$ and $Z$ can
each be empty) such that $Y$ contains the most preferred candidates of $v$, $v$ is weakly
increasing along $X$ and $v$ is weakly decreasing along $Z$.
\end{definition}

Notice that the above definition extends single-plateauedness to allow
for multiple indifference plateaus on either side of the peak. So for votes with ties,
possibly single-peaked preferences model when voters have weakly
increasing and then weakly decreasing or only weakly increasing/decreasing preferences
along an axis.

Another generalization of single-peaked preferences for votes with ties was introduced by
Cantala~\shortcite{can:j:single-peaked-outside-option}: the model of single-peaked preferences with outside options. %
When preferences satisfy this restriction with respect to a given axis, each voter has a segment
of the axis where they have single-peaked preferences and candidates appearing outside of
this segment on the axis are strictly less preferred and the voter is tied between
them. Similar to how single-plateaued preferences extend the standard single-peaked
model to
allow voters to state multiple most preferred candidates, single-peaked preferences with
outside options extends the standard model to allow voters to state multiple
least preferred candidates.

\begin{definition}
Given a preference profile $V$ of weak orders over a set of candidates $C$,
$V$ is single-peaked with outside options with respect to a total ordering of the candidates $L$ (an axis)
if for each voter $v \in V$, $L$ can be split into five segments $O_1$, $X$, $Y$, $Z$, and $O_2$
($O_1$, $X$, $Z$, and $O_2$ can each be empty)
such that $Y$ contains only the most preferred candidate of $v$, $v$ is strictly
increasing along $X$ and $v$ is strictly decreasing along $Z$,
for all candidates $a \in X \cup Y \cup Z$ and $b \in O_1 \cup O_2$, $v$ states $a > b$,
and for all candidates $x,y \in O_1 \cup O_2$, $v$ states $x \sim y$.
\end{definition}

Menon and Larson~\shortcite{men-lar:c:sp-manipulation-partial} state that the model of single-peakedness %
for top orders that they use
is similar to single-peaked preferences with outside options. %
It is clear from their paper that for top orders these models are the same.
So for the remainder of the paper we will refer to the model used
by Menon and Larson~\shortcite{men-lar:c:sp-manipulation-partial} as
``single-peaked preferences with outside options for top orders.'' %

\subsection{Social Choice Properties}\label{sec:res:general}

We now state some general observations %
on single-peaked preferences with ties, including
how the
models relate to each other, as well as their social-choice properties.

\begin{figure*}[t]
  \centering
  \subfloat[Relationship for weak orders]{\includegraphics[scale=0.75]{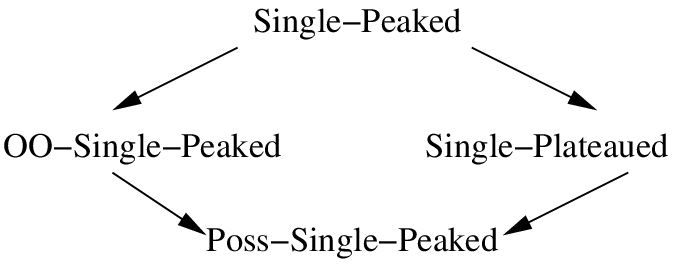}}\hfill
  \subfloat[Relationship for top orders]{\includegraphics[scale=0.75]{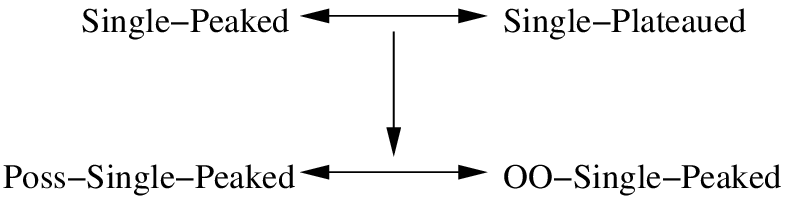}}\hfill
  \subfloat[Relationship for bottom orders]{\includegraphics[scale=0.75]{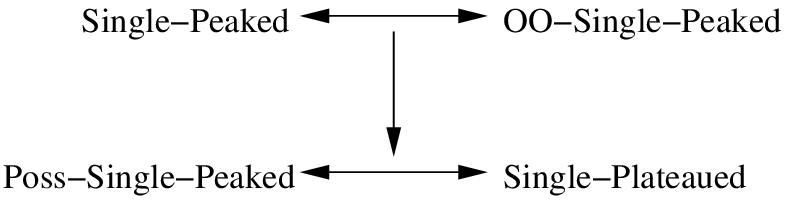}}
\caption{Relationships between the four models of single-peaked preferences for different
types of votes with ties, where $A \to B$ indicates that a preference profile satisfying model $A$ also satisfies model $B$.
OO-Single-Peaked and Poss-Single-Peaked refer to Single-Peaked with
Outside Options and Possibly Single-Peaked respectively.}
\label{fig:relation}
\end{figure*}

It is easy to see that for total-order preferences each of the four models of single-peakedness with ties
that we consider are equivalent.
In Figure~\ref{fig:relation} we show how each model relates for weak orders, top orders, and bottom orders.

Given a preference profile of votes it is natural to ask how to determine if an axis exists
such that the profile satisfies one of the above restrictions. This is referred to as the consistency problem for a restriction.
Bartholdi and Trick~\shortcite{bar-tri:j:stable-matching-from-psychological-model}
showed that single-peaked consistency for total orders can be determined in
polynomial time (i.e., is in \p),
and Fitzsimmons~\shortcite{fit:c:weak-order-sp} extended this result to show that single-peaked, single-plateaued, and possibly single-peaked consistency
for weak orders is in \p. %
This leaves the consistency problem for single-peaked preferences with outside options, which we will now show to be in \p.

It is easy to see that given a preference profile of top orders, it is single-peaked with
outside options if and only if it is possibly single-peaked.
So it is immediate from the result by Lackner~\shortcite{lac:c:incomplete-sp-aaai} that shows that possibly single-peaked consistency for top orders
is in \p, %
that the consistency problem for single-peaked preferences with outside options for top orders is
in \p. For weak orders
the construction used to show that single-peaked consistency for weak orders is in \p\ 
by Fitzsimmons~\shortcite{fit:c:weak-order-sp} can be adapted to hold for the case of
single-peaked preferences with outside options for weak orders, so this consistency problem is also in \p.

\begin{theorem}
  Given a preference profile $V$ of weak orders it can be determined in polynomial time if there
  exists an axis $L$ such that $V$ is single-peaked with outside options
  with respect to $L$.
\end{theorem}

One of the most well-known and desirable properties of an election with single-peaked preferences is that
there exists a transitive majority relation~\cite{bla:j:rationale-of-group-decision-making}.
A majority relation is transitive if for all distinct candidates $a,b,c \in C$ if $a > b$ and $b > c$
by majority then $a > c$ by majority. When the majority relation is transitive then candidates that beat-or-tie
every other candidate by majority exist and they are denoted the weak Condorcet winners.
This also holds when the voters have single-plateaued preferences~\cite{bla:b:polsci:committees-elections}.

It is not the case that the majority relation is transitive when preferences in an election are
single-peaked with outside options or are possibly single-peaked.
Fitzsimmons~\shortcite{fit:c:weak-order-sp} points out that for possibly single-peakedness this 
was implicitly shown in an entry of Table~9.1 in
Fishburn~\shortcite{fis:b:theory}, where a preference profile of top orders that violates
single-peaked preferences was
described that does not have a transitive majority relation, and this profile is possibly single-peaked (and single-peaked with outside options).
Cantala also provides an example of a profile that is single-peaked with outside options that does
not have a weak Condorcet winner, and so does not have a transitive majority relation~\cite{can:j:single-peaked-outside-option}.

\section{Computational Results}\label{sec:results}

For total-order preferences, 
weighted manipulation for any fixed number of candidates is known to be \npc\ for
every scoring rule that is not isomorphic to plurality or triviality~\cite{hem-hem:j:dichotomy}.
For single-peaked total orders, Faliszewski et al.~\shortcite{fal-hem-hem-rot:j:single-peaked-preferences}
completely characterized the complexity of
weighted manipulation for
3-candidate scoring rules %
and this result was generalized
by Brandt et al.~\shortcite{bra-bri-hem-hem:j:sp2} for any fixed number of candidates. %
These results both showed that the complexity of weighted manipulation for scoring rules
often decreases when voters have single-peaked total orders.

For top-order preferences,
dichotomy theorems for 3-candidate weighted manipulation for scoring rules using round down,
min, and average were 
shown by Menon and Larson for the general case~\cite{men-lar:t:manipulation-partial} and
for the case of single-peaked preferences with outside options for top orders~\cite{men-lar:c:sp-manipulation-partial}.
Menon and Larson~\shortcite{men-lar:c:sp-manipulation-partial} found that, counterintuitively, the complexity of weighted manipulation for
single-peaked preferences with outside options %
often increases when moving from total orders to top
orders. %
We also mention that for the scoring-rule extension
max, we earlier found that the complexity of 3-candidate weighted manipulation for Borda increases for
possibly single-peaked preferences when moving from total orders to
top orders~\cite{fit-hem:c:voting-with-ties}.

We show that for the standard model of single-peakedness and for single-plateauedness
that the complexity of weighted manipulation for $m$-candidate scoring rules using max, min, round down,
and average does not increase when moving from total orders to top orders, bottom orders, or weak orders.
The following results are 
close analogs to the case for single-peaked total orders due to
Brandt et al.~\shortcite{bra-bri-hem-hem:j:sp2}.

\begin{lemma}\label{lem:sp-first}
  If $p$ can be made a winner by a manipulation of top-order, bottom-order,
or weak-order votes that are single-peaked, single-plateaued, possibly single-peaked, or single-peaked with outside options for a scoring rule using max, min, round
down, or average, then $p$ can be made a winner by a manipulation (of the same type)
in which all manipulators rank $p$ uniquely first.
\end{lemma}

\begin{proofs}
  Suppose that a manipulator has a vote of the form $(A_{\rm wo} > P > B_1 > \cdots > B_\ell)$
  where $A_{\rm wo}$ is a weak order over $A$ (where $A$ can be empty), $\ell \ge 0$, each of $P,B_1,\ldots,B_\ell$ are nonempty groups of
  tied candidates, and $p \in P$.
  We have the following two cases.

  \begin{description}
   \item[Case 1:] If all of the candidates in $P$ appear on the same side of the candidates in $A$ with respect to the axis (or $A$ is empty), then
   we can change the vote of the manipulator to $p > {(P-\{p\})}_{\rm to} > A_{\rm to} > B_1 > \cdots > B_{\ell}$
   where ${(P-\{p\})}_{\rm to}$ and $A_{\rm to}$ are total orders such that the vote satisfies the given notion of single-peakedness with ties with respect
   to the axis.
   \item[Case 2:] Otherwise, we let $P'$ be the candidates to the left of $A$ and $P''$ be the candidates to the right of $A$ with respect to the axis.
   Without loss of generality let $p \in P'$. We can then change the vote of the manipulator 
   to $p > {(P'-\{p\})}_{\rm to} > A_{\rm to} > {P''}_{\rm to} > B_1 > \cdots > B_{\ell}$.
  \end{description}

  Notice that in all of the above cases, for all candidates $c \in C-\{p\}$ that $\score{p} - \score{c}$ does not decrease,
  regardless of which scoring-rule extension is used.
  So if $p$ is a winner with the initial vote, $p$ is still a winner in the vote that
  ranks $p$ uniquely first.~\end{proofs}

\begin{theorem}\label{thm:noninc}
Let $\alpha = \langle\alpha_1, \alpha_2, \ldots, \alpha_m\rangle$
be a scoring vector. If $\alpha$-CWCM is in P for single-peaked total
orders, then
$\alpha$-CWCM is in P for single-peaked and single-plateaued
top orders, bottom orders, and weak orders for all our scoring rule extensions.
\end{theorem}

\begin{proofs}
If the societal axis has $p$ in the leftmost or rightmost position,
then, by Lemma~\ref{lem:sp-first}, we can assume that all manipulators rank
$p$ uniquely first. 
There is exactly one such single-peaked or single-plateaued vote with ties,
namely to rank $p$ first and
strictly rank the remaining candidates according to their position on the axis. 

So, let the societal axis be
$a_{m_1} L \cdots L a_1 L p L b_1 L \cdots L b_{m_2}$.

We first consider the \p\ cases for single-peaked total orders described
in Lemma~6.6 of Brandt et al.~\shortcite{bra-bri-hem-hem:j:sp2}
where for all $i,j > 1$ such that $i+j \le m+1$ it holds that
$(\alpha_1 - \alpha_i)(\alpha_1-\alpha_j) \le (\alpha_i - \alpha_{i+1})(\alpha_j - \alpha_{j+1})$.

The crucial observation in the proof of this lemma is that for every set
of single-peaked total order
votes $S$, we are in one of the following two cases:
\begin{enumerate}
\item For all $i$, $1 \leq i \leq m_1$,
 $\scoresub{S}{a_i} \leq \scoresub{S}{p}$.
\item For all $i$, $1 \leq i \leq m_2$,
 $\scoresub{S}{b_i} \leq \scoresub{S}{p}$.
\end{enumerate}
We will show that this is also the case for single-peaked and single-plateaued
orders with ties. This then implies that the optimal vote for all
manipulators is
$p > b_1 > \cdots > b_{m_2} > a_1 > \cdots > a_{m_1}$ (in case~1) 
or
$p > a_1 > \cdots > a_{m_1} > b_1 > \cdots > b_{m_2}$ (in case~2).

For the sake of contradiction let $S$ be a collection of nonmanipulators and $i_1$ and $i_2$
be integers such that $1 < i_1 \le m_1 +1$ and $1 < i_2 \le m_2 + 1$ such that
$\scoresub{S}{a_{i_1}} > \scoresub{S}{p}$ and $\scoresub{S}{b_{i_2}} > \scoresub{S}{p}$. 

We now create an ${\widehat S}$ such that $\scoresub{{\widehat S}}{a_{i_1}} > \scoresub{ {\widehat S} }{p}$ and
$\scoresub{ {\widehat S} }{b_{i_2}} > \scoresub{ {\widehat S}}{p}$ and the argument in the proof
of Lemma~6.6 in Brandt et al.~\shortcite{bra-bri-hem-hem:j:sp2} still holds.

\begin{itemize}
  \item First, remove from $S$ all voters that have $p$ tied for first. Notice that if we remove all votes with $p$
    tied for first from $S$ then $\score{p}$ does not increase and neither $\score{a_{i_1}}$ nor
    $\score{b_{i_2}}$ decreases.
    So we can assume that $p$ is never tied for first in a vote (although this does not mean that
    other candidates cannot be tied for first in a vote when we are in the single-plateaued case).

  \item We know that $p$ is not tied for first (part of an indifference plateau), so $p$ is tied with
  at most one other candidate in each vote. So in every vote break the tie against $p$. Notice that
  this can lower only the score of $p$ regardless of whether we are using the min, max, round down, or average
  scoring-rule extension.
\end{itemize}

Let $\ell_i$ be the total weight of the voters in ${\widehat S}$ that rank some candidate in
  $\{a_1, \ldots, a_{m_1}\}$ first and $p$ in the position of $\alpha_i$ (in every extension other than
  round down this means that $p$ is ranked $i$th; for round down we mean that $p$ is ranked in the
  $\alpha_i$ group (i.e., $p$ is followed by $m-i$ groups).
Using a similar argument to the proof of Lemma~6.6 from Brandt et al.~\shortcite{bra-bri-hem-hem:j:sp2}
with ${\widehat S}$ as described above we can reach a contradiction. %

We now present the remaining \p\ cases, which can be seen as the
with-ties analog of the \p\ cases of Lemma~6.7
from Brandt et al.~\shortcite{bra-bri-hem-hem:j:sp2}.
  We examine each of the cases individually.
For convenience, we normalize the scoring vector so that $\alpha_m = 0$.

  If $\alpha_2 = 0$ then clearly %
  the optimal action for the manipulators is to vote a total order with $p$ uniquely first and the
  remaining candidates ranked arbitrarily such that the vote is single-peaked.

 If $\alpha_1 = \alpha_{\lfloor \frac{m-1}{2} \rfloor + 2}$, then,
as pointed out in~\cite{bra-bri-hem-hem:j:sp2}, Lemma 6.6 applies, which has
been handled above.

The last P-time case is that 
$\alpha_1 \le 2\alpha_2$, $\alpha_1 > \alpha_2 > 0$, and
$\alpha_2 = \alpha_{m-1}$.

  Notice that when the votes are single-peaked or when they are single-plateaued then
  $a_{m_1}$ and $b_{m_2}$ are the only candidates that can occur last in a vote, either uniquely
  or tied (except for the case where all candidates are tied, but we can easily ignore these votes
  since they do not change the relative scores between candidates).

  Since we know from Lemma~\ref{lem:sp-first} that all of the manipulators can put $p$ strictly ranked
  first, we can have the following manipulator votes.
  \begin{itemize}
    \item $p > \cdots > a$.
    \item $p > \cdots > b$.
    \item $p > \cdots > a \sim b$.
  \end{itemize}
  
  Notice that for all votes with $p$ first that $\score{p} \ge \alpha_2$.

  For all votes, $\score{a_{m_1}} = 0$ or $\score{b_{m_2}} = 0$, or $a_{m_1}$ and $b_{m_2}$ are tied for last, in which case
  $\score{a_{m_1}} + \score{b_{m_2}} \le 2\alpha_2$ (the worst case occurs with max where they each receive
  a score of $\alpha_{m-1}$).

  So, $\scoresub{S}{a_{m_1}} + \scoresub{S}{b_{m_2}} \le 2\scoresub{S}{p}$.
If $\scoresub{S}{a_{m_1}} \le \scoresub{S}{p}$,
$p > a_1 > \cdots > a_{m_1} > b_1 > \cdots > b_{m_2}$ is the optimal manipulator vote.
Otherwise, $\scoresub{S}{b_{m_2}} \le \scoresub{S}{p}$ and
$p > b_1 > \cdots > b_{m_2} > a_1 > \cdots > a_{m_1}$ is the optimal manipulator vote.

We have proven that the above cases hold for single-peaked and for single-plateaued weak orders, and it is clear
that the case for single-peaked bottom orders follows from the case of single-peaked total orders (since they
are equivalent) and it is also clear
how to adapt the above arguments to hold for single-peaked top orders, and for single-plateaued top
orders and bottom orders.~\end{proofs}

The most surprising result in the work by Menon and Larson~\shortcite{men-lar:c:sp-manipulation-partial}
was that the complexity of 3-candidate CWCM %
for elimination 
veto for single-peaked preferences with outside options for top orders using min is NP-complete, whereas %
for single-peaked total orders~\cite{men-lar:c:sp-manipulation-partial} and
even for total orders
in the general case~\cite{col-tea:cNEW:manipulation} it is in \p.

\begin{theorem}\cite{col-tea:cNEW:manipulation}
  $m$-candidate elimination veto CWCM for total orders is in \p\ in the unique-winner model.
\end{theorem}

\begin{theorem}\cite{men-lar:c:sp-manipulation-partial}
  $m$-candidate elimination veto CWCM for single-peaked total orders is in \p\ in the unique-winner model.
\end{theorem}

\begin{theorem}\cite{men-lar:c:sp-manipulation-partial}
 3-candidate elimination veto CWCM for top orders that are single-peaked with outside options is \np-complete in the unique-winner model.
\end{theorem}

Menon and Larson~\shortcite{men-lar:c:sp-manipulation-partial} state this case
as a counterexample to the conjecture by Faliszewski et al.~\shortcite{fal-hem-hem-rot:j:single-peaked-preferences}, which states
that the complexity for a natural election system will not increase when moving from the general case to the single-peaked case.
(Though Menon and Larson do qualify that the conjecture from Faliszewski et al.\ 
concerned
total orders.) However, for
the standard model of single-peaked preferences and for single-plateaued
preferences, elimination veto CWCM for
top orders, bottom orders, and weak orders using min is in \p\ for any fixed number of candidates,
thus the counterexample crucially relies on using a nonstandard definition.

\begin{theorem}
 $m$-candidate elimination veto CWCM for single-peaked and for single-plateaued %
 top orders, bottom orders, and weak orders using min is in \p\ in the unique-winner model.
\end{theorem}

\begin{proofs}
The proof of this theorem follows from a similar argument to the proof of the theorem for
single-peaked total orders from Menon and Larson~\shortcite{men-lar:c:sp-manipulation-partial}. This is
because that proof follows from the fact that for total orders
the candidate eliminated after each round is on the leftmost or
rightmost location of the axis, and the reverse of an elimination order is single-peaked with respect
to the axis. It is easy to see that both of these statements also hold for single-peaked votes with ties, since the only
candidates that can be vetoed in each round are still the leftmost and rightmost candidates on the axis.
For single-plateaued preferences it is possible for voters with an indifference plateaued to veto more
than two candidates (after all of the candidates ranked below their indifference plateau have been eliminated),
but it is easy to see that these votes do not affect which candidate is eliminated since all of the
remaining candidates are vetoed by such voters. It is also clear that Lemma~12 from
Coleman and Teague~\cite{col-tea:cNEW:manipulation}
still holds which states that if there exists a collection of votes that can induce an elimination order,
then this elimination order can be induced by all of the manipulators voting the reverse of the elimination order.
So since we know that the reverse of an elimination order is single-peaked with respect to the axis, and that
there are only polynomially many possible elimination orders with $p$ first, the manipulators simply try each
elimination order with $p$ first.

Since the elimination order is always a total order, the above argument clearly holds for single-peaked
and single-plateaued top orders, bottom orders, and weak orders.~\end{proofs}

It is known that 3-candidate Copeland$^\alpha$  CWCM for all rational $\alpha \in [0,1)$ is NP-complete for
total orders in the nonunique-winner
model~\cite{fal-hem-sch:c:copeland-ties-matter},
and when $\alpha = 1$ (also known as Llull) CWCM is in \p\ for
$m \le 4$~\cite{fal-hem-sch:c:copeland-ties-matter,fal-hem-sch:c:llull4}
and the cases for $m \ge 5$ remain open.
Fitzsimmons and Hemaspaandra~\shortcite{fit-hem:c:voting-with-ties} showed that the NP-completeness of
the 3-candidate case holds
for top orders, bottom orders, and weak orders, and
Menon and Larson~\shortcite{men-lar:t:manipulation-partial}
independently showed the top-order case.
Fitzsimmons and Hemaspaandra~\shortcite{fit-hem:c:voting-with-ties}
also showed that 3-candidate Llull CWCM is in \p\ for top orders, bottom orders,
and weak orders. %

Recall that
weak Condorcet winners always exist when preferences are single-peaked with
ties, and when they are single-plateaued. So the results that Llull CWCM for single-peaked total
orders is in \p\ from Brandt et al.~\shortcite{bra-bri-hem-hem:j:sp2} also holds for the case of
single-peaked and single-plateaued top orders, bottom orders, and weak orders.

Copeland$^\alpha$ for $\alpha \in [0,1)$ was shown to be in \p\ by
Yang~\shortcite{yan:c:manip-sp-width} for single-peaked total
orders.

\begin{theorem}\cite{yan:c:manip-sp-width}
\label{t:yang}
Copeland$^\alpha$ CWCM for $\alpha \in [0,1)$ is in \p\ 
for single-peaked total orders.
\end{theorem}

In contrast, for top orders that are single-peaked with outside options,
it is NP-complete even for three candidates~\cite{men-lar:c:sp-manipulation-partial}.

\begin{theorem}\cite{men-lar:c:sp-manipulation-partial}
3-candidate Copeland$^\alpha$ CWCM for $\alpha \in [0,1)$ is
\np-complete for top orders that are single-peaked with outside options.
\end{theorem}

For single-peaked and single-plateaued 
weak orders, bottom orders, and top orders we again
inherit the behavior of single-peaked total orders.

\begin{theorem}
Copeland$^\alpha$ CWCM for $\alpha \in [0,1)$ is in \p\ for
single-peaked and single-plateaued top orders, bottom orders, and weak orders.
\end{theorem}

\begin{proofs}
Let $L$ be our axis.
Consider a set of nonmanipulators with single-plateaued weak orders. 
Replace each nonmanipulator $v$ of weight $w$ 
with two nonmanipulators $v_1$ and $v_2$ of weight $w$.
The first nonmanipulator breaks the ties in the vote in increasing
order of $L$ and the second nonmanipulator breaks the ties
in the vote in decreasing order of $L$,
i.e., if $a \sim_v b$ and $a L b$, then $a >_{v_1} b$ and $b >_{v_2} a$.
Note that $v_1$ and $v_2$ are single-peaked total orders and that
the weighted majority graph induced by the 
nonmanipulators after replacement can be obtained from the
weighted majority graph induced by the original nonmanipulators by
multiplying each weight in the graph by 2.
When we also multiply the manipulator weights by 2, we have an equivalent
Copeland$^\alpha$ CWCM problem, where all nonmanipulators are
single-peaked total orders and all manipulators have even weight.

Suppose $p$ can be made a winner by having the manipulators cast
single-plateaued votes with ties. Now replace each manipulator
of weight $2w$ by two weight-$w$ manipulators. The first manipulator
breaks the ties in the vote in increasing order of $L$ and the second
manipulator breaks the ties in the vote in decreasing order of $L$.
Now the replaced manipulator votes are single-peaked total orders and $p$
is still a winner.  We need the following fact from the proof of
Theorem~\ref{t:yang}: \emph{if $p$ can be made a winner 
in the single-peaked total order case, then $p$ can be made a
winner by having all manipulators cast the same P-time
computable vote.} It follows
that $p$ can be made a winner by having all replaced manipulators
cast the same single-peaked total order vote. But then $p$ can
be made a winner by having all original manipulators cast the 
same single-peaked total order vote. Since this vote is P-time 
computable, it follows that
Copeland$^\alpha$ CWCM for $\alpha \in [0,1)$ is in \p\ for
single-plateaued weak orders, and this also holds for single-peaked weak orders
since every single-peaked profile of weak
orders is also single-plateaued.

It is clear to see that similar arguments hold for single-peaked top orders,
and for single-plateaued top orders and bottom orders.
The case for single-peaked bottom orders follows from the case for single-peaked
total orders.~\end{proofs}

\subsection{Societal-Axis Results}\label{sec:splitaxis}

Recall from Lemma~\ref{lem:sp-first}
that for all our single-peaked models and all our scoring rule extensions,
we can assume that all manipulators rank $p$ uniquely first.
When given an axis where the preferred candidate is in the leftmost or
rightmost location, there is exactly one single-peaked total order
vote that puts
$p$ first, namely, $p$ followed by
the remaining candidates on the axis in order.
This is also the case for single-peaked and single-plateaued orders
with ties, since $p$ is ranked uniquely first and 
no two candidates can be tied on the same side of the peak.
It follows that the weighted manipulation problems for scoring rules
for single-peaked total orders and single-peaked and single-plateaued orders
with ties are in P for axes where $p$ is in the leftmost/rightmost position. %
However, this is not the case when preferences with ties are single-peaked with outside options or
possibly single-peaked. In Case~1 of the
proof of Theorem~1 in the work by Menon and Larson~\shortcite{men-lar:c:sp-manipulation-partial},
an axis of $p L a L b$
is used to show \np-hardness for their model. %

It is attractive to conjecture that for single-peaked and
single-plateaued preferences, the less symmetrical (with respect to $p$)
the axis is,
the easier the complexity of manipulation,
but surprisingly
this turns out to not be the case, even for total orders.

For the theorem stated below let $m_1$ and $m_2$ denote the number of candidates to the left and to the
right on the axis with respect to the %
preferred candidate of the manipulators.

\begin{theorem}
For single-peaked total orders, single-peaked weak orders, and 
single-plateaued weak orders, 
$\langle 4,3,2,0,0 \rangle$ CWCM %
is in \p\ for $m_1 = m_2 = 2$ and \np-complete for
$m_1 = 1$ and $m_2 = 3$ for all our scoring rule extensions.
\end{theorem}

\begin{proofs}
$\mathbf{m_1 = m_2 = 2}$:
Careful inspection of the proof of Lemma~6.6
from Brandt et al.~\shortcite{bra-bri-hem-hem:j:sp2} proves the
following refined version of that lemma that takes the axis into account.

\begin{lemma}\label{l:sp2p}
Let $\alpha = \langle \alpha_1, \alpha_2, \cdots, \alpha_m \rangle$ be
a scoring rule.
If for all $i,j$ such that $1 < i \le m_1 + 1$ and $1 < j \le m_2 + 1$,
it holds that
    $$(\alpha_1 - \alpha_i)(\alpha_1 - \alpha_j) \le (\alpha_i - \alpha_{i+1})(\alpha_j - \alpha_{j+1})$$
    then $\alpha$-CWCM %
for single-peaked total orders for this axis is in \p.
  \end{lemma}

Note that Lemma~\ref{l:sp2p} applies in our case, since for all
$i,j$ such that $1 < i \leq 3$ and $1 < j \leq 3$, the following hold.

\begin{tabular}{ccccc}
$i$ & $j$ & $(\alpha_1 - \alpha_i) (\alpha_1 - \alpha_j)$ & & 
$(\alpha_i - \alpha_{i+1}) (\alpha_j - \alpha_{j+1})$ \\
2 & 2 &
$(4-3)(4-3)=1$ & $\leq$ & $(3-2)(3-2)=1$\\
2 & 3 &
$(4-3)(4-2)=2$ & $\leq$ & $(3-2)(2-0)=2$\\
3 & 2 & 
$(4-2)(4-3)=2$ & $\leq$ & $(2-0)(3-2)=2$\\
3 & 3 &
$(4-2)(4-2)=4$ & $\leq$ & $(2-0)(2-0)=4$
\end{tabular}

It follows that $\langle 4,3,2,0,0 \rangle$ CWCM is in P for single-peaked
total orders when %
$m_1 = m_2 = 2$.
It follows from the arguments in the proof of Theorem~\ref{thm:noninc} %
that $\langle 4,3,2,0,0 \rangle$ CWCM is also in P for single-peaked and
single-plateaued orders with ties for $m_1 = m_2 = 2$.

To get NP-completeness for the single-peaked and single-plateaued 
weak order cases for all our scoring rule extensions, we need to 
modify the proof for single-peaked total orders.
Careful inspection of the proof of Lemma~6.4 from Brandt
et al.~\shortcite{bra-bri-hem-hem:j:sp2}
allows us to extract the following 
reduction from Partition to $\langle 4,3,2,0,0 \rangle$
CWCM %
for total
orders that are single-peaked with respect to
axis $a_1 L p L b_1 L b_2 L b_3$.

Given an instance of Partition,
a collection of $t$ positive integers $k_1, \ldots, k_t$
such that $\sum_{i=1}^{t} k_i = 2K$, we are asking whether there
exists a partition of this collection, i.e., whether there
exists a subcollection that sums to $K$.

The set $S$ of nonmanipulators consists of
\begin{itemize}
  \item One weight $20K$ voter voting: $a_1 > p > b_1 > b_2 > b_3$.
  \item One weight $8K$ voter voting: $b_1 > b_2 > b_3 > p > a_1$.
\end{itemize}

The weights of the manipulators are $4k_1, 4k_2, \ldots, 4k_t$.

If there exists a partition, then
let the set $T$ of manipulators vote the following.
\begin{itemize}
  \item Coalition of weight $4 K$ manipulators vote
$(p > a_1 > b_1 > b_2 > b_3)$.
  \item Coalition of weight $4 K$ manipulators vote
$(p > b_1 > b_2 > b_3 < a_1)$.
\end{itemize}

Then $\score{p} = \score{a_1} = \score{b_1} = 92K$, and the scores of the other
candidates are lower. For the converse, note that the optimal
total order votes for the
manipulators are $(p > a_1 > b_1 > b_2 > b_3)$ and
$(p > b_1 > b_2 > b_3 > a_1)$.
Simple calculation shows that in order to make $p$ a winner, exactly half
of the manipulator weight needs to
vote $(p > a_1 > b_1 > b_2 > b_3)$.
This corresponds to a partition.

For the single-peaked and single-plateaued cases for votes with ties, note that
by Lemma~\ref{lem:sp-first}, we can assume that all manipulators rank $p$ uniquely first.
We will look at each of our scoring extensions: max, min, average,
and round down.

For max, if the manipulators can make $p$ a winner, we break the
ties of the manipulators such that the resulting votes
become total single-peaked orders. $p$ will still be a winner in the
resulting election. This implies
that the construction above also gives NP-completeness for single-peaked 
votes with ties using max.

For round down, the construction above also works.
Suppose $p$ can be made a winner. Again, we assume that all manipulators
rank $p$ uniquely first. Now replace all votes with ties
(i.e., $(p > b_1 \sim a_1 > b_2 > b_3)$, 
$(p > b_1 > b_2 \sim a_1 > b_3)$, and $(p > b_1 > b_2 > b_3 \sim a_3)$) by
$(p > b_1 > b_2 > b_3 > a_1)$.  Note that $\score{p} - \score{a_1}$
and $\score{p} - \score{b_1}$ do not decrease, and so $p$ is still a
winner of the resulting election, in which all manipulators vote a
single-peaked total order. 
This implies
that the construction above also gives NP-completeness for single-peaked 
votes with ties using round down.

For average, the construction above also works, though the proof
needs a little more work.
Suppose $p$ can be made a winner. Again, we assume that all manipulators
rank $p$ uniquely first. So, $\score{p} = 92K$. The optimal manipulator votes
are: $(p > a_1 > b_1 > b_2 > b_3)$, $(p > b_1 > b_2 > b_3 > a_1)$,
and $(p > a_1 \sim b_1 > b_2 > b_3)$. Assume that these are the only manipulator
votes and let $W_a$, $W_b$, and $W_{ab}$ be the total manipulator weight voting 
each of these three votes, respectively.
Since $p$ is a winner, $\scoresub{T}{a_1} \leq 12K$ and
$\scoresub{T}{b_1} \leq 20K$.
It follows that
$3W_a + 2.5W_{ab} \leq 12K$, $2W_a + 3W_b + 2.5 W_{ab} \leq 20K$, and
$W_a + W_b + W_{ab} = 8K$. Standard calculation shows that this
implies that $W_a = 4K$, $W_b = 4K$ and $W_{ab} = 0$. Thus the
weights of the voters voting $(p > a_1 > b_1 > b_2 > b_3)$
correspond to a partition.

For min, the optimal votes for the manipulators are 
$(p > a_1 \sim b_1 > b_2 > b_3)$ and $(p > b_1 > b_2 > b_3 > a_1)$.
In order to make
the argument from the total order case work, we need to adjust the weights of the
nonmanipulators a bit. Namely, we let the set $S$ of nonmanipulators consist of
\begin{itemize}
  \item One weight $24K$ voter voting: $a_1 > p > b_1 > b_2 > b_3$.
  \item One weight $9K$ voter voting: $b_1 > b_2 > b_3 > p > a_1$.
\end{itemize}
The weights of the manipulators are unchanged, i.e., $4k_1, \ldots, 4k_t$.
Note that if half the manipulator weight votes
$(p > a_1 \sim b_1 > b_2 > b_3)$ and remaining
manipulators vote $(p > b_1 > b_2 > b_3 > a_1)$, then
$\score{p} = (24 \cdot 3 + 8 \cdot 4) K = 104K$,
$\score{a_1} = (24 \cdot 4 + 4 \cdot 2) K = 104K$, and
$\score{b_1} = (24 \cdot 2 + 9 \cdot 4 + 4 \cdot 2 + 4 \cdot 3) K =
(48 + 36 + 8 + 12) K = 104K$. The scores of $b_2$ and $b_3$ are lower.
Similarly to the previous cases, if $p$ can be made a winner then
standard calculations show that exactly half of the manipulator
weight votes $(p > a_1 \sim b_1 > b_2 > b_3)$, which
corresponds to a partition.~\end{proofs}

\section{Related Work}\label{sec:related}
The work by Menon and Larson~\shortcite{men-lar:c:sp-manipulation-partial} on the complexity of manipulation and bribery
for single-peaked preferences with outside options for top orders is the most closely related to this paper.
For manipulation, they show that for single-peaked preferences with outside options the complexity
often increases when moving from total orders to top orders.
They additionally considered a notion of nearly single-peakedness. %
We instead study the complexity of weighted manipulation for the standard model of single-peaked preferences with ties
and for single-plateaued preferences with ties.

The focus of our paper is on the computational aspects of models of single-peaked preferences with ties.
These models can also be compared based on which social-choice properties that they have, such as the
guarantee of a weak Condorcet winner. Barber{\'a}~\shortcite{bar:j:indifference-domain} compares
such properties of the models of single-peaked, single-plateaued, and single-peaked with
outside options for votes with ties.

Since single-peakedness is a strong restriction on preferences, in real-world scenarios it is likely that
voters may only have nearly single-peaked preferences, where different
distance measures to a single-peaked profile are considered. Both the computational
complexity of different manipulative attacks~\cite{fal-hem-hem:j:nearly-sp,erd-lac-pfa:c:k-app-sp} and detecting %
when a given profile is nearly single-peaked~\cite{erd-lac-pfa:c:nearly-sp,bre-che-woe:j:nicely-structured-nearby} have been considered.
An important computational problem for single-peakedness is determining the
axis given a preference profile, this is known as its consistency problem.
Single-peaked consistency for total orders was first shown to be in \p\ by Bartholdi and
Trick~\shortcite{bar-tri:j:stable-matching-from-psychological-model}. 
Doignon and Falmagne~\shortcite{doi-fal:j:unidimensional-unfolding} and Escoffier, Lang, and
{\"O}zt{\"u}rk~\shortcite{esc-lan-ozt:c:single-peaked-consistency} independently found
faster direct algorithms. Lackner~\shortcite{lac:c:incomplete-sp-aaai}
proved that possibly single-peaked consistency for top orders
is in \p\ (and for local weak orders and partial orders is \np-complete), and
Fitzsimmons~\shortcite{fit:c:weak-order-sp}
later showed that single-peaked, single-plateaued, and possibly single-peaked consistency
for weak orders is in \p.

\section{Conclusions}\label{sec:conclusion}

The standard model of single-peakedness is naturally defined for votes with ties, but different
extensions have been considered. %
In contrast to recent work that studies the models of possibly single-peaked and
single-peaked preferences with outside options %
and finds an anomalous increase in complexity compared to the tie-free case,
we find that for scoring rules and other important natural systems, 
the complexity of weighted manipulation does not increase when moving from total orders to votes with ties in the %
standard single-peaked and in the single-plateaued cases.
Single-peaked and single-plateaued preferences
for votes with ties also retain the important social-choice property of the
existence of weak Condorcet winners.
This is not to say that possibly single-peaked and single-peaked
preferences with outside options are without merit, since
they both model easily understood structure in preferences.

\smallskip

\noindent
{\bf Acknowledgments:}
We thank the anonymous referees for their helpful comments.
This work was supported in part by NSF grant no.\ CCF-1101452 and a NSF
Graduate Research Fellowship under NSF grant no.\ DGE-1102937.

\end{document}